\documentclass[final,unnumberedheadings,letterpaper]{aipproc}

\layoutstyle{6x9}
\newcommand{\lya}{Ly$\alpha$}

\usepackage[]{pslatex}
\usepackage[]{natbib}
\usepackage[]{amssymb}
\usepackage[]{myusepack}

\begin{document}
        
%%%%%%%%%%%%%%%%%%%%%%%%%%%%%%%%%%%%%%%%%%%%
%% FRONTMATTER
%%%%%%%%%%%%%%%%%%%%%%%%%%%%%%%%%%%%%%%%%%%%

\title[Essential LyC]{Essential observations of the Lyman continuum }

%                \texttt{http://www.aip.org/pacs/index.html}}

\author{Stephan R. McCandliss}{
  address={Department of Physics and Astronomy, The Johns Hopkins University, Baltimore, MD  21218}, altaddress={stephan@pha.jhu.edu}
}

\begin{abstract}
Concurrent observations of Lyman continuum (LyC) and Lyman $\alpha$ (\lya) emission escaping from star-forming systems at low redshift are essential to understanding the physics of reionization at high redshift ($z \ga$ 6).  Some have suggested reionization is dominated by numerous small galaxies with LyC escape fractions $f_e \sim$ 10\%, while others suggest mini-quasars with higher $f_e$ might also play a role.  At  $z >$ 3, direct observation of LyC leakage becomes progressively more improbable due to the increase of intervening Ly limit systems, leaving \lya\ as the primary diagnostic available to the James Webb Space Telescope for exploring the epoch of reionization.  If a quantitative relationship between escaping LyC and \lya\ emission can be established at low $z$, then the diagnostic power of \lya\ as a LyC proxy at high $z$  can be fully realized.  Past efforts to detect $f_e $ near $z \approx$ 3 have been fruitful but observations at low redshift have been less so.  We discuss the sensitivity requirements for detecting LyC leak in the far- and near-UV as a function of redshift 0.02 $< z \lesssim$ 3 and $f_e \ge$ 0.01 as estimated from UV luminosity functions.  UV observations are essential to understanding of the physics of LyC escape and the ultimate goal of identifying the source(s) responsible for reionization.
\end{abstract}

\maketitle

%%%%%%%%%%%%%%%%%%%%%%%%%%%%%%%%%%%%%%%%%%%%
%% MAINMATTER
%%%%%%%%%%%%%%%%%%%%%%%%%%%%%%%%%%%%%%%%%%%%

\subsection{Introduction}
%\noindent{\bf Introduction --}
We know most of the universe is ionized and approximately when it happened.  Sloan Digital Sky Survey spectra of luminous high redshift quasars have black \ion{H}{1} Gunn-Peterson troughs, indicating a mean \ion{H}{1} fraction of $\gtrsim 10^{-3}$ at  $z \ge $ 6.4 \citep{Fan:2006} when the age of universe was $\approx$ 1 Gyrs.  An earlier constraint is provided by the polarization of the microwave background on large angular scales, seen by the Wilkinson Microwave Anistropy Probe, which is consistent with an ionization fraction $\sim$ unity at $z \approx$ 11 when the universe was $\approx$ 365 Myrs old \citep{Spergel:2007}.  It is suspected the reionization epoch may have started even earlier.    

Reionization occurs when the number of ionizing photons emitted within a recombination time exceeds the number of neutral hydrogen atoms.  The duration of the reionization epoch depends on the initial mass function of the first ionizing sources, their intrinsic photoionization rate ($Q$), the baryon clumping factor ($C\equiv <\rho^2>/<\rho>^2$), and the fraction of ionizing photons ($f_e$) that somehow escape into the intergalactic medium (IGM)  \citep{Madau:1999}.   LyC escape is a poorly constrained parameter \citep{Fan:2006}, the arbitrary choice of which can alter conclusions regarding the nature and duty cycle of the sources responsible for initiating and sustaining  reionization \citep[e.g.][]{Gnedin:2000}.  Here I broadly review the subject and then discuss how UV observations are the only means to directly examine the physical conditions that favor escape of the LyC.  Thus, future UV observations are essential to our understanding the physics behind reionization.

\subsection{Essential role LyC escape }
%\noindent{\bf Essential role of LyC escape}
LyC escape plays an essential role in the formation of structure.  The escape fraction of ionizing photons from galaxies has been identified as the greatest single uncertainty in estimating the evolution of the intensity of the metagalactic ionizing background (MIB) over time \citep{Heckman:2001}. The MIB influences the ionization state of the IGM at all epochs and may be responsible for hiding a non-trivial fraction of the baryons in the universe \citep[c.f.][]{Tripp:2008, Danforth:2008}.  Ionizing radiation produced by star-forming galaxies is ultimately related to the rate of metal production by stellar nucleosynthesis \citep{Gnedin:1997}, providing a limit to the rate at which supernovae and stellar winds can feedback chemicals, mechanical energy and radiation into the IGM.

Ionizing radiation regulates the collapse of baryons on local and global scales \citep{Ricotti:2008}.  Photoelectrons produced in the ionization process provide positive radiative feedback on star formation by promoting the formation of H$_{2}$, as catalyzed by the H$^{-}$.  H$_{2}$ is a crucial coolant for collapse at high-z.  Negative feedback occurs when photoionization heating raises the temperature and inhibits stellar collapse by increasing the Jeans mass.  Photodissociation of H$_{2}$ is another form of negative radiative feedback mediated by both Lyman-Werner photons in the "{\it FUSE}" bandpass and LyC photons \citep{McCandliss:2007}.  Whether ionizing radiation has a positive or negative effect on a collapsing body depends on the density, the strength of the radiation field, the source lifetime  and  $f_e$ \citep{Ricotti:2008, Ciardi:2008}.

\subsection{Sources of LyC} 
%\noindent{\bf Sources of LyC}
The essential question is, how did the universe come to be reionized and how long did it take?  Current thinking posits that LyC escape from the smallest galaxies powers reionization at $z \approx$ 6, since quasars are too few in number to sustain reionization \citep{Madau:1999, Bouwens:2008, Yan:2004}.  However, this conclusion depends on the slope at the faint end of the galaxy luminosity function (LF, $\alpha \le $-1.7), the clumping factor (20 $ < C < $45) and $f_e \sim$ 0.1 -- 0.2.

The slope at the faint end of the quasar LF near $z \sim$ 6 is poorly constrained, so it is not clear if quasars can produce ionizing radiation on par with that required from galaxies to sustain reionization \citep{Fan:2004}. Nevertheless, there are indications that the initiation of reionization above $z = $ 7 may require a "hard" spectral energy distribution (SED) more characteristic of quasars, or perhaps an $f_e > $ 0.2 that increases with redshift \citep{Bolton:2007, Meiksin:2005}.  It has been suggested there may not be enough galaxies early on to initiate reionization \citep{Bolton:2007} and that mini-quasars might be involved \citep{Madau:2004}.

It is now accepted that black holes reside in the nuclei of most if not all quiescent galaxies \citep{Ferrarese:2005, Soltan:1982}, so it is perhaps simplistic to characterize reionization as a process caused by either quasars with $f_e$ = 1 or star-forming galaxies with $f_e <$  1.  Some fraction of quasars do exhibit a break at the Lyman edge \citep{Kriss:1997} likely due to obscuration by host galaxies, resulting in a softer SED \citep{Shull:2004}.  The central engines of active galactic nuclei (AGN) come in a variety of masses and have intermittent duty cycles, so the effects of previous AGN activity within an apparently dormant galactic environment may reduce for a time the local \ion{H}{1} density and allow the LyC produced by stars to escape.

\subsection{Escape and the \lya\ proxy}

%\noindent{\bf Detection of LyC escape and the \lya\ proxy}
In the coming decade the James Webb Space Telescope (JWST) will seek to identify the source(s) responsible for initiating and sustaining reionization. 
\citet{Inoue:2008} have shown that for $z >$ 3 the likelihood of detecting LyC escape from star-forming galaxies becomes increasingly improbable, due to a progressive increase, with increasing redshift, in the number density of intervening Lyman limit systems having column densities of 10$^{17} < N_{HI} < $ 2 $\times$ 10$^{20}$ cm$^{-2}$.  JWST will have to rely upon observations of \lya\ escape as a proxy for LyC escape.  Unfortunately there is no guarantee that such a proxy relationship exists, because escaping \lya\ photons are created by recombining electrons freed by the LyC photons that do not escape \citep{Stiavelli:2004} ([\lya\ $\approx (2/3)Qf_{Ly\alpha}(1-f_e)exp{-\tau}$]).  It is thus essential to test the proxy hypothesis at $ z \la$ 3 by obtaining simultaneous observations of LyC and \lya.   

\begin{figure}[h]   
\includegraphics[height=.27\textheight]{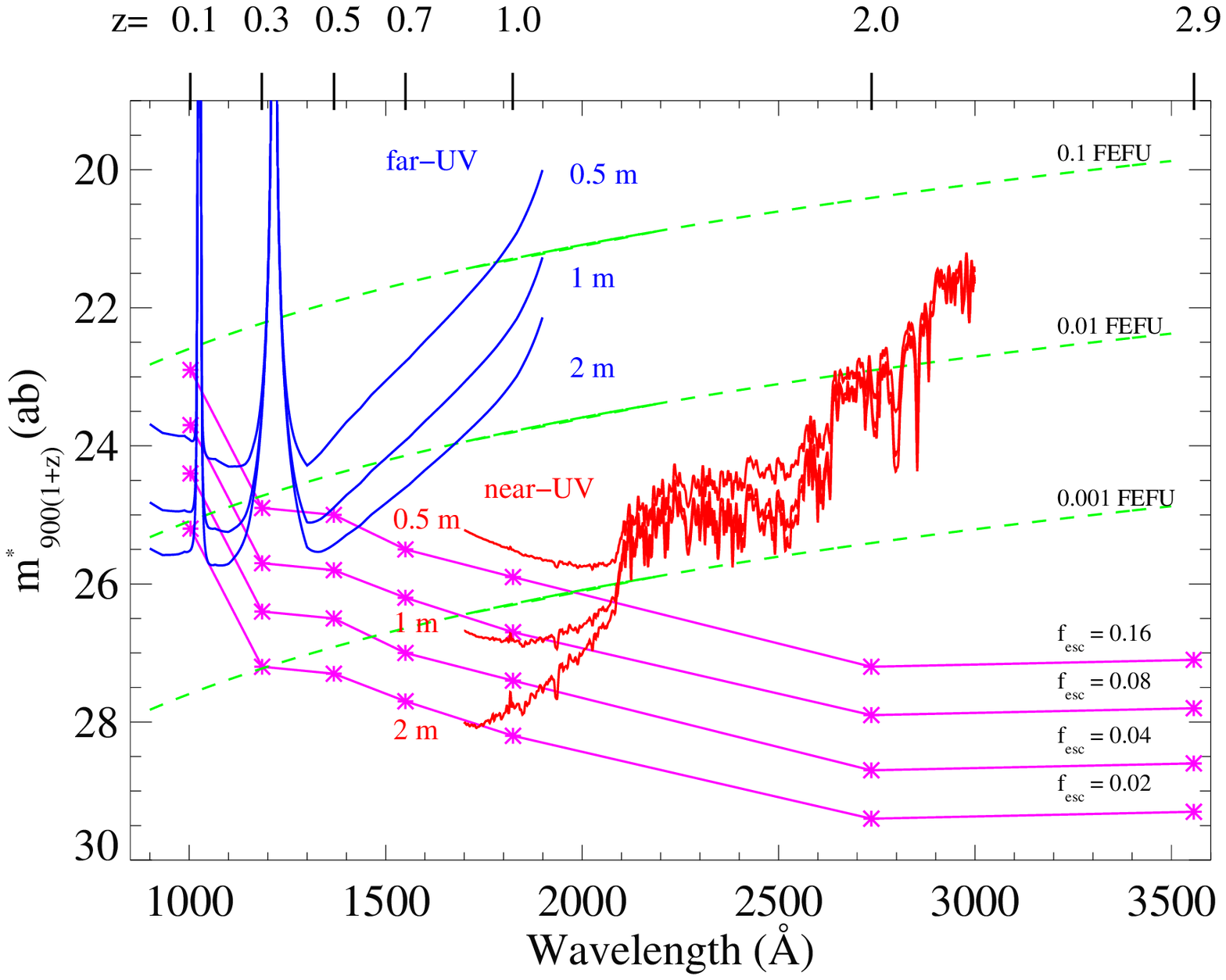}\includegraphics[height=.3\textheight]{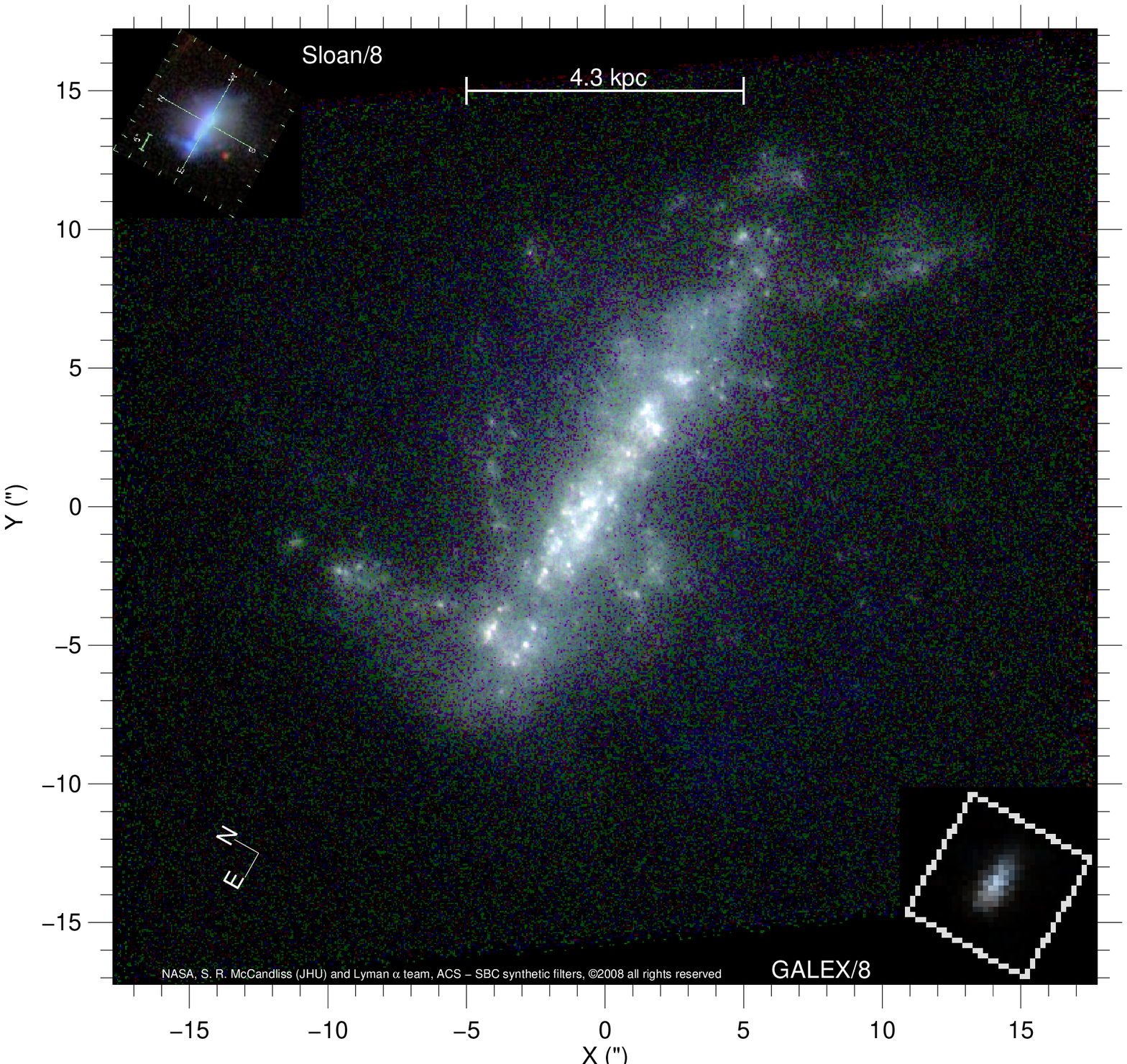}
\caption{LEFT -- Apparent magnitude of the LyC emitted by $m*_{900(1+z)}$ galaxies as a function of redshift and escape fraction. Background limits  for high efficiency UV spectro/telescopes with 0.5 m, 1 m, and 2 m apertures are overplotted. Flux densities ($f_{\lambda}$) in FEFU $\equiv$ 10$^{-15}$ (ergs cm$^{-2}$ s$^{-1}$ \AA$^{-1}$) are plotted with long-dashs. The background is  dominated by geo-coronal airglow in the far-UV and zodiacal light in the near-UV, entering a slit of 36$\arcsec\ \times $17$\arcsec$. RIGHT -- ACS/SBC synthetic filter image of Mrk 66.  Blue shows regions of diffuse \lya\ leakage.}
\end{figure}

%\subsection{How do LyC and \lya\ escape?}
%\paragraph{\bf How do LyC and \lya\ escape?}

Reionization appears to require LyC leakage from galaxies with $f_e \sim 0.1$, but how LyC and \lya\ escape from galaxies is somewhat mysterious.  Most star-forming galaxies have mean \ion{H}{1} columns greater than damped \lya\ systems (DLA have $N_{HI} \ge $ 2 $\times$ 10$^{20}$ cm$^{-2}$),  so the optical depth at the Lyman edge is, $\tau_{\lambda<912}$ = $N_{HI}$ 6.3 x 10$^{-18}$ $(\lambda/912)^3$ = 1260 $(\lambda/912)^3$, while at the line core of \lya\ the optical depth is, $\tau_{Ly\alpha}$ = $N_{HI}$ 6.3 x 10$^{-14}$ = 1.26 $\times$ 10$^{7}$, for $V_{dop}$ = 12 km s$^{-1}$.   Escape from such large mean optical depths requires that the interstellar medium be highly inhomogeneous.  LyC escape is thought to result from  galaxy porosity; low neutral density, high ionization voids or chimneys created by supernovae or the integrated winds from stellar clusters \citep{Fujita:2003}.  The escape of \lya\ is aided by velocity gradients and resonance scattering in a multi-phase media \citep{Hansen:2006}.

\subsection{Previous efforts to detect LyC escape \\[-.15in]}
%\paragraph{Previous efforts to detect LyC escape}
Observations to date indicate that LyC escape from star-forming galaxies is difficult to detect.  The results are rather mixed.  There are no reported observations of LyC escape for $z \ge $3.4.  \citet{Steidel:2001} report excess LyC in a composite spectral stack of  29 Lyman Break Galaxies (LBG) with a mean $<z>$ = 3.4, while \citet{Shapley:2006} report direct LyC leak detections in 2 LBG at $z \sim$ 3.  Both these reports  tentatively estimate that  $f_e$ is high, $\sim$ 0.5.  In contrast, \citet{Fernandez-Soto:2003} found only an upper limit of $f_e <$  0.04 in Hubble Deep Field imaging of 27 galaxies with spectroscopic redshifts 1.9 $< z <$ 3.5.   \citet{Iwata:2008} report 17 detections, using narrowband imaging, which they estimate to have a median  $f_e$  $  > $ 0.04, assuming isotropic escape.  \citet{Siana:2007} report an image stack limit for the relative $f_e <$ 0.08  of 21 sub $L^*$\footnote{$L_{uv}^*$ is the characteristic UV luminosity of a galaxy luminosity distribution function.} galaxies in the HDF-N and HUDF in the redshift range  1.1 $< z < $1.5.  At low, $z \sim$ 0.02, there exist only controversial detections and upper limits with $f_e$ $<$ 1 - 57\% \citep[c.f.][]{Leitherer:1995, Hurwitz:1997, Deharveng:2001, Heckman:2001, Bergvall:2006, Grimes:2007}.  \\[-.45in]

\subsection{LyC detection requirements and open questions \\[-.15in]}
%\paragraph{LyC detection requirements and open questions}
These previous efforts suggest that future instruments capable of determining escape limits to 1\% for $L_{uv}^*$ galaxies will provide a comprehensive assessment of whether LyC can escape from star-forming systems for 0.02 $< z \la$ 3.   This is challenging, as evidenced by the left panel of Figure~1  from \citet{McCandliss:2008}, which shows estimated sensitivity requirements for detecting LyC leak in $L_{uv}^*$ galaxies down to $f_e \ge$ 0.01 as a function of redshift 0.02 $< z \lesssim$ 3, computed from rest frame UV LFs \citep{Arnouts:2005}.  To reach $f_e \approx$ 0.01  will require apertures well in excess of 2 m. Observations beyond $z >$ 1.3 will be especially challenging, because of zodiacal light.   

The question of detection hinges on how $f_e$ changes as a function of luminosity. The usual picture is that low mass galaxies have the highest  $f_e$ \citep{Ricotti:2000}  however, a recent study by \citet{Gnedin:2008} finds that  $f_e$  declines steeply with decreasing mass galaxies.   A related question concerns the existence of analogs at low-z to the faint galaxies at high-z responsible for reionization.  \citet{Ricotti:2005} have suggested that as the MIB grew in strength it shut off star-formation in the low density halos before they became metal enriched.  If these systems survived to the present day and are now beginning to form stars, then attempts to determine their $f_e$ should be a high priority.    Such questions will require wide field surveys to properly address. The findings will be important regardless of the outcome.  If star-forming galaxies are found with  $f_e \ga$ 0.1 then they become plausible sources of reionization.  If not, then new physics will be required to explain reionization \citep{Gnedin:2008}, provided quasar number counts remain low.

The new complement of UV instruments scheduled to be installed on {\it HST} is well suited for initial studies. The UV sensitive CCD's of WFC3 can acquire ultra deep fields as short as 2180 \AA\ and search for "not quite" LBGs.  COS can be used to look for leak at $z >$ 0.2 and it has the advantage of providing simultaneous observation of LyC and \lya.   Programs started with {\it FUSE} aimed at redshifts of 0.02 $< z <$ 0.04, but never completed because of its untimely demise, could become viable if the far-UV channel on COS proves to be sensitive.  ACS/SBC images acquired in support \lya\ observations of {\it FUSE} LyC candidates (right panel Figure~1) provide a high precision finding chart for differential LyC leak studies in select regions of low-z galaxies.  Future high-grasp UV telescopes sensitive enough to detect LyC leak will easily detect the cosmic web of low surface brightness \lya\ emission; an unambiguous beacon for emerging complexity \citep{Furlanetto:2003}. \\[-.5in]

%%%%%%%%%%%%%%%%%%%%%%%%%%%%%%%%%%%%%%%%%%%%
%% Sample figure:
%%
%% The option [height=...] scales the picture to the given height,
%% without it it would be printed at its nominal size
%%%%%%%%%%%%%%%%%%%%%%%%%%%%%%%%%%%%%%%%%%%%

%%%%%%%%%%%%%%%%%%%%%%%%%%%%%%%%%%%%%%%%%%%%
%% SAMPLE TABLE
%%
%% Shows the use of \tablehead and \tablenote
%% macros
%%%%%%%%%%%%%%%%%%%%%%%%%%%%%%%%%%%%%%%%%%%%

%%%%%%%%%%%%%%%%%%%%%%%%%%%%%%%%%%%%%%%%%%%%%%%%
%% BACKMATTER
%%%%%%%%%%%%%%%%%%%%%%%%%%%%%%%%%%%%%%%%%%%%%%%%

\subsection{Acknowledgments \\[-.15in]}
I wish to thank all the members of Projects Lyman and Balmer for their support.  NASA grants to JHU NNG04WC03G and NNX08AM68G support this work. \\[-.4in]



\bibliographystyle{aipproc}   % if natbib is available

\bibliography{essential}

\IfFileExists{\jobname.bbl}{}
 {\typeout{}
  \typeout{******************************************}
  \typeout{** Please run "bibtex \jobname" to optain}
  \typeout{** the bibliography and then re-run LaTeX}
  \typeout{** twice to fix the references!}
  \typeout{******************************************}
  \typeout{}
 }

\end{document}